\documentclass[referee,a4paper,12pt,traditabstract]{swsc} 

\usepackage{graphicx}
\usepackage{txfonts}
\usepackage{subfigure}
\usepackage{epstopdf}
\usepackage{lineno}
\usepackage[authoryear,round]{natbib}
\usepackage[backref]{hyperref}
\usepackage{url}
\usepackage{braket}
\usepackage{booktabs}
\usepackage{upgreek}

\hypersetup{colorlinks=true,citecolor=cyan,urlcolor=cyan,linkcolor=blue}

\begin{document}

	 \title{Improvements on coronal hole detection in SDO/AIA images using supervised classification}		  
	
   \titlerunning{Improvements on coronal hole detection in SDO/AIA images} 

   \authorrunning{M.A.~Reiss et al.}
	
   \author{Martin A.~Reiss\inst{1}, Stefan J.~Hofmeister\inst{1}, Ruben De Visscher\inst{2}, Manuela Temmer\inst{1}, Astrid M.~Veronig\inst{1}, V\'{e}ronique Delouille\inst{2}, Benjamin Mampaey\inst{2}, Helmut Ahammer\inst{3}}

   \institute{University of Graz, IGAM-Kanzelh\"ohe Observatory, NAWI Graz, Graz, Austria\\
              \email{\href{mailto:martin.reiss@uni-graz.at}{martin.reiss@uni-graz.at}} \and
							Royal Observatory of Belgium, Brussels, Belgium \\
							\email{\href{mailto:veronique.delouille@oma.be}{veronique.delouille@oma.be}} \and
							Medical University of Graz, Institute of Biophysics, Graz, Austria}

  \date{Received 19 December 2014 / Accepted 20 June 2015}
 
  \abstract{
	We demonstrate the use of machine learning algorithms in combination with segmentation techniques in order to distinguish coronal holes and filaments in 
	SDO/AIA EUV images of the Sun. Based on two coronal hole detection techniques (intensity-based thresholding, SPoCA), we prepared data sets of manually labeled 
	coronal hole and filament channel regions present on the Sun during the time range 2011 - 2013. By mapping the extracted regions from EUV observations onto 
	HMI line-of-sight magnetograms we also include their magnetic characteristics. We computed shape measures from the segmented binary maps as well as first 
	order and second order texture statistics from the segmented regions in the EUV images and magnetograms. These attributes were used for data mining 
	investigations to identify the most performant rule to differentiate between coronal holes and filament channels. We applied several classifiers, namely 
	Support Vector Machine, Linear Support Vector Machine, Decision Tree, and Random Forest and found that all classification rules achieve good results in general, with linear 
	SVM providing the best performances (with a true skill statistic of $\approx 0.90$). Additional information from magnetic field data systematically improves the performance across all four classifiers for the SPoCA detection. Since the calculation is inexpensive in computing time, this approach is well suited for applications on real-time data. This study demonstrates how a machine learning approach may help improve upon an unsupervised feature extraction method.
   }       

   \maketitle

\section{Introduction}
	Coronal holes play an important role in geomagnetic storm activity \citep{tsurutani06} and are the dominant contributors 
		to space weather disturbances at times of quiet solar activity. The term coronal hole is commonly associated with regions of 
		one dominant magnetic polarity with rapidly expanding open magnetic field lines along which solar wind particles escape into 
		interplanetary space \citep{gosling99, cranmer09}. Coronal holes are the source regions of the high speed solar wind streams (HSSs). 
		In combination with the Sun's rotation, they shape the solar wind distribution in interplanetary space. In Extreme 
		Ultraviolet (EUV) and X-ray images of the Sun, coronal holes are visible as dark areas in the solar corona due 
		to their lower temperature and electron density compared to the ambient coronal plasma \citep{munro72}. 

		The detection of coronal holes purely from their low intensity in solar EUV images is a challenging task since filament channels also appear as 
		dark coronal features. Filament channels are usually interpreted in terms of the weakly twisted flux rope model, having a magnetic field which is 
		dominated by the axial component. Dense prominence material is located in the dip of the helical windings leading to the elongated dark structures observed 
		at a similar dark intensity level as coronal holes \citep{mackay10, detoma11}. 
		
		In the past, coronal holes have mostly been identified and tracked by experienced observers. There have been recent attempts to automate the process 
		for the identification and detection of coronal holes \citep{barra2007datafusion,kirk09,krista09,rotter12}. Those detection techniques are mainly based on differences 
		in intensity compared to the ambient corona. 
		
		The extraction of coronal holes is of interest for various applications in astrophysics, e.g. the forecast of HSSs. The areas of coronal holes extracted at 
		the solar meridian reveal a correlation with the solar wind speed measured $\sim$4 days later in-situ at 1~AU \citep{vrsnak07, verbanac11, rotter12, rotter15}. 
		This relation enables us to forecast HSSs based on coronal hole observations. Thus, it is important to improve the coronal hole detection method, in order to 
		exclude filaments that may erroneously be identified as a coronal hole region in EUV images. 

		In this work, Solar Dynamic Observatory (SDO) Atmospheric Imaging Assembly (AIA) 19.3~nm images \citep{lemen12} and SDO/ Helioseismic and Magnetic Imager 
		(HMI) magnetograms \citep{scherrer12} are used in order to analyse the properties of coronal holes and filament channels during the period 2011 January 1 till 
		2013 December 31. We first obtain low intensity regions from SDO/AIA 19.3~nm images by means of two coronal hole detection techniques: one using
		intensity thresholding \citep{rotter12}, and a second, called SPoCA \citep{verbeeck14}, based on fuzzy clustering of intensity values. Comparing these detections 
		with H$\upalpha$ filtergrams from Kanzelh\"ohe Observatory (Austria) in which filaments can be clearly identified \citep{poetzi14}, we label these regions as 
		either \lq filament' or \lq coronal hole'. The detected regions were mapped onto the corresponding HMI line-of-sight magnetograms. This allows us to calculate a 
		set of attributes that takes into account EUV intensity and binary shape information as well as the magnetic field structure for each of the regions. 
			
		Our contribution in this paper is two-fold. First, we devise decisive coronal feature attributes. We focus on the magnetic flux imbalance, first and second order 
		image statistics and shape measures since: (i) coronal holes are expected to be characteristic regions of one dominant magnetic polarity whereas filament channels 
		are regions of closed magnetic field lines along a magnetic field reversal line; (ii) the physical properties of filament channels are partly reflected in their 
		geometric characteristics. Filament channels are, in contrast to coronal holes, in most cases observable as elongated structures that extend in a particular direction.
		
		\begin{figure}%
		\begin{center}
		\includegraphics[width=0.999\columnwidth]{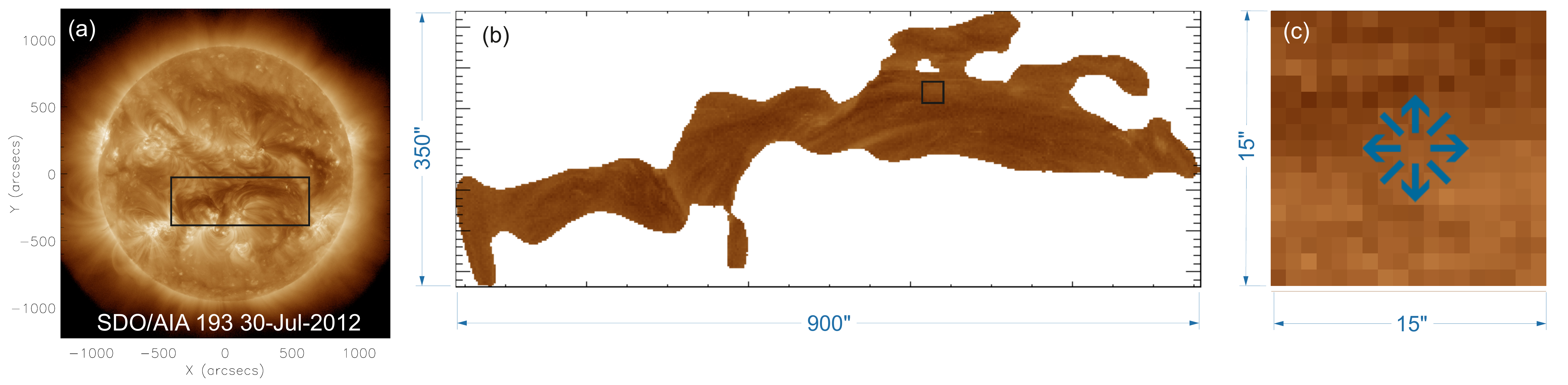}
		\caption{Analysis of the detected low intensity regions. (a) AIA 19.3~nm image from 30-Jul-2012 showing a filament channel; (b) Detected filament channel 
		from EUV image based on intensity thresholding; (c) Illustration of the spatial relationship of pixel values contained in the filament channel.} 
		\label{fig:figure1}%
		\end{center}
		\end{figure}
		
		Second, we insert the computed attributes into various supervised classification schemes in order to design the most suitable decision rule for 
		a differentiation in real-time. We tested and compared through cross-validation four commonly used classifiers, namely Support Vector Machine (SVM), Linear SVM, 
		Decision Tree, and Random Forest. To evaluate the performance of different classifiers we made use of the Hanssen \& Kuipers discriminant, also known as 
		true skill statistics (TSS) \citep{hanssen65,bloomfield12}, as a quantitative measure. This paper is structured in the following way: In 
		Section~\ref{sec:datapreparation} the preparation of the labeled data sets is explained. A description of the proposed attributes is given in 
		Section~\ref{sec:methods} and the applied supervised classification schemes are presented in Section~\ref{S:classificationproblem}. The results are given 
		in Section~\ref{sec:results} and their implications are discussed in Section~\ref{sec:discussion}.

\section{Data preparation}\label{sec:datapreparation}
	
		Two different coronal hole segmentation techniques (intensity-based thresholding \citep{rotter12, rotter15}, SPoCA \citep{verbeeck14}) were applied in order 
		to prepare two data sets of low intensity regions during the period 2011 January 1 till 2013 December 31. Based on these data we created training sets to 
		design the most suitable decision rule for a differentiation of coronal hole and filament channel regions. 
	
	\subsection{Coronal hole feature extraction}\label{sec:featureextraction}

	\subsubsection{Intensity-based thresholding}\label{sec:segmentation1}	
		An intensity-based thresholding technique was used to detect low intensity regions in SDO/AIA 19.3~nm images. Based on the intensity distribution in the full-disk EUV images 
		we apply a threshold value of $0.38 \times \mbox{(median intensity value)}$ for areas within $\pm60^\circ$ longitude and latitude. Due to the optically thin emission 
		from neighboring coronal structures, we used an additional multiplication factor of 1.6 for areas outside the $\pm60^\circ$ longitude and latitude window.
		With this refinement, polar coronal holes are also well detected although they have a smaller contrast. From this we created binary maps, which were post-processed 
		by using morphological methods (erosion, dilation).	The effects of these operators on binary maps are to erode (or shrink) the boundaries of regions, removing all 
		small anomalies (erosion) and to gradually expand the boundaries of regions to fill small holes in-between (dilation). A $16 \times 16$ pixel kernel was used to remove 
		small artefacts and fill small gaps in the extracted regions.

		\subsubsection{Fuzzy clustering}\label{sec:segmentation2}
		
		The SPoCA-suite~\citep{verbeeck14} is a set of segmentation procedures that allows decomposition of an EUV image into regions of similar intensity, typically 
		active regions, coronal holes, and quiet sun. 
		 In the framework of the Feature Finding Team project (FFT; \cite{2012SoPh..275...79M}), the SDO Event Detection System 
		(SDO EDS; \cite{2012SoPh..275...67H}) runs the SPoCA-suite to extract coronal hole information from AIA images in the 19.3~nm passband, and upload the entries every four hours to 
		the Heliophysics Events Knowledgebase~\citep{2012SoPh..275...67H}. These entries are searchable through the graphical interface iSolSearch, the ontology software 
		package of IDL Solarsoft, and the JHelioviewer visualization tool~\citep{jhelio}.
		
		The SPoCA extraction method for coronal holes relies on the Fuzzy C-Means algorithm (FCM)~\citep{Bezdek:1981}. It clusters the pixel's intensity values into four 
		classes through the minimization of the intra-cluster variance. Such minimization leads to an iterative algorithm, where at each iteration every pixel is 
		assigned a membership value between 0 and 1 to each one of the class, then the class centers are computed using these memberships. These steps are repeated until
		convergence of the class centers. The segmented map is obtained by attributing a pixel to the class for which it has the maximum final membership value. The coronal hole 
		map corresponds to the class whose center has the lowest intensity value. Various preprocessing steps are performed before applying FCM algorithm: Images are 
		calibrated using the IDL Solarsoft \verb|aia_prep| routine, and intensities are normalized by their median values. We correct for the limb brightening effect by 
		fitting a smooth transition function and inverting it. Finally, a square-root transform is applied on the images. Indeed,~\cite{ansc} showed that for Poisson distributed 
		data a square-root transform induces exact asymptotic normality and stabilizes the variance.  This is especially useful for the extraction of low-intensity regions which 
		are affected by Poisson noise. 

		Once the segmented maps are obtained, some post-processing is needed to extract individual low intensity regions. Elements with a radius smaller than 6 arcsec are 
		removed and neighbouring connected components within a 64 arcsec distance are aggregated. Coronal holes candidates having an area smaller than $3000$ arcsec$^2$ are 
		discarded. A study of low intensity feature detected by SPoCA during the month of January 2011 revealed that coronal hole candidates detected for more than three 
		consecutive days exhibit the expected magnetic properties characteristic of unipolar regions, which was not the case for shorter lived regions. Setting up a threshold 
		on lifetime is thus a valuable tool to eliminate spurious dark regions. Hence only low-intensity regions which live longer than three days are included in the final 
		coronal hole maps used here and in the SDO EDS pipeline (we refer to~\cite{verbeeck14} for more details).

	\subsection{Labeled datasets}

    The extracted regions were manually labeled as either \lq filament' or \lq coronal hole' regions by visually inspecting H$\upalpha$ filtergrams from Kanzelh\"ohe 
		Observatory (Austria). We selected only regions greater than $1900$ arcsec$^2$ for which Kanzelh\"ohe H$\upalpha$ data were available and that were located close to the 
		central meridian ($\pm30^\circ$ in longitude/latitude). If at the position of an extracted low intensity region a filament was clearly observable in the corresponding H$\upalpha$ 
		filtergram (to take into account the evolution of filament material we checked a time span of three days), the structure was labeled as \lq filament'. Otherwise, 
		the structure was labeled as \lq coronal hole'. This procedure was performed for both segmentation methods once per day, 
		resulting in two training sets: one containing 349 coronal holes and 61 filaments for the intensity-based thresholding method and the second with 252 coronal holes and 
		46 filaments for the SpoCA algorithm. In addition we mapped the extracted regions onto HMI line-of-sight magnetograms to quantify the magnetic flux characteristics 
		in the extracted regions. Note that both segmentation methods have different capabilities for extracting low-intensity regions, which explains the dissimilarity in the 
		number of CH candidates detected. In future works, we plan on comparing more exhaustively these two methods.

\section{Proposed attributes} \label{sec:methods}
		On the prepared training datasets we investigate shape measures, magnetic flux properties and first and second order image statistics for the calculation 
		of characteristic attributes. Shape measures were used to characterize the shape of the detected low intensity regions. The calculated shape measures also allow us to 
		reduce the object pixel configuration contained in a binary map of a structure to a single scalar number. With the help of first order and second order image 
		statistics \citep{haralick73, weyn00, ahammer08} we focus on the overall pixel value distribution and the spatial relation between pixel values contained in AIA 
		EUV images and HMI line-of-sight magnetograms. Second order statistics, originally applied on grey scale images, was adapted to open value ranges (i.e. including
		negative data values) in order to calculate textural features for intensity and magnetic field configurations. This enables us to characterize the intrinsic 
		texture information contained in the extracted structures in AIA 19.3~nm and line-of-sight magnetograms via the computation of a set of textural features. 
		A description of the proposed methods is outlined in the following subsections. 
		
	\subsection{Shape Measures}\label{sec:shapemeasures}
	
		While many shape descriptors exist, there exists no generally accepted definition in literature. In order to investigate irregular shapes of coronal holes and 
		filament channels in detail, two alternative shape measurements were developed.
		\begin{description}
		\item[\bf{Symmetry analysis}]
		
		Shapes are intuitively described with the terms symmetric or asymmetric. We measure geometrical symmetry properties
		with the following technique: After the application of discrete geometrical transformations like rotation, reflection and a 
		composition of both, the relative overlap in percentage is calculated.
		
		\item[\bf{Direction-dependent shape analysis}]
		
		Shapes can also be presented in form of a function, e.g. the average number of neighbours in each direction is representative 
		for the relevant shape information of an object. For a detailed description of the used shape measures we refer to \cite{reiss14}.

		\end{description}
		
	\subsection{Magnetic Flux Imbalance}\label{sec:fluximbalance}
		We characterize the magnetic flux imbalance in the detected features by calculating the following two attributes from the line-of-sight
		magnetograms:
		\begin{equation}
		R_1 = \frac{n_{+}}{n_{-}} \,, 
		\label{eq:}
		\end{equation} 
		i.e. the ratio of the number of positive ($n_{+}$) and negative ($n_{-}$) pixel values, and
		\begin{equation}
		R_2 = 2 \ \left|\frac{1}{2} - \frac{\Phi_+}{\Phi_+ + \left|\Phi_- \right|} \right| \,,
		\label{eq:}
		\end{equation}
		with $\Phi_+$ being the total positive magnetic flux in the segmented feature, and $\Phi_-$ being the total negative flux. $R_2$ quantifies 
		the imbalance between the total positive and the total magnetic flux within the extracted regions. In the ideal case, this value would be 
		$1$ for coronal holes, as regions of a unique polarity ($\Phi_+=0$ or $\Phi_-=0$), and $0$ for filament channels ($\Phi_+=\Phi_-$), as magnetically 
		bipolar regions orientated along a polarity inversion line. 

\subsection{First Order Image Statistics}\label{sec:1storderstatistics}
		Pixel values contained in the extracted regions were rounded to the nearest integer value. The probability $P(i)$ for the occurrence of pixels 
		with integer value $i$ is defined as
		\begin{equation}
		P(i) = \frac{n(i)}{N},
		\label{eq:}
		\end{equation}
		where $n(i)$ is defined as the number of pixels with the actual pixel value $i$ and $N$ is the total number of image pixels. 
		The following standard first order attributes were used:
		\begin{equation}
		\mbox{Mean: } \ \mu = \sum_{i}{i \ P(i)},
		\label{eq:}
		\end{equation}
		\begin{equation}
		\mbox{Variance: } \ \sigma^2 = \sum_{i}{(i - \mu)^2 P(i)},
		\label{eq:}
		\end{equation}
		\begin{equation}
		\mbox{Standard Deviation (Contrast): } \ C_1 = \sqrt{\sigma^2},
		\label{eq:}
		\end{equation}
		\begin{equation}
		\mbox{Energy: } \ E_1 = \sum_{i}{P(i)^2},
		\label{eq:}
		\end{equation}
		\begin{equation}
		\mbox{Entropy: } \ S_1 = \sum_{i}{-P(i) \log (P(i))}.
		\label{eq:}
		\end{equation}
		Variance $\sigma^2$ and contrast $C_1$ are associated with the width of the pixel value distribution. High variance or contrast indicate large 
		pixel value differences. Energy $E_1$ is high for unevenly distributed pixel values independent of the designated pixel value itself. High entropy 
		$S_1$ values reflect the degree of information content within the pixel value distribution. Low entropy values correspond to coherent pixel values 
		with low information content. High values reflect highly disordered image values with a high amount of intrinsic information content.
				
	\subsection{Second Order Image Statistics}\label{sec:2ndorderstatistics}
		Second order statistics \citep{haralick73, weyn00, ahammer08} are utilized for the calculation of textural features of AIA and HMI images providing 
		information about the spatial arrangement of pixel values. A set of matrices $n_{\vec{\phi, d}}(i,j)$ count how many 
		times a given combination of pixel values $i$ and $j$ occur in a particular spatial arrangement. The probabilities of co-occurrences of pixel values 
		can then be represented with the so called co-occurrence matrix $C(i,j)$. The basic idea how spatial information is contained in low intensity regions 
		is illustrated in Figure~\ref{fig:figure1}. Textural features allow the description of textural information with a set of scalar numbers. 
	
		\subsubsection{Co-Occurrence Matrix} 
		Analogous to the probability $P(i)$ for the occurrence of pixels with value $i$, we define $P_{\phi, d}(i,j)$ indicating the probability for 
		the co-occurrence of pixel value $i$ and pixel value $j$ in a given distance $d$ and direction $\phi$. $n_{\vec{\phi, d}}(i,j)$ is the total 
		number of co-occurrences of pixel value $i$ with pixel value $j$ within the distance $d$ and direction $\phi$. The matrix $P_{\phi, d}(i,j)$ 
		is defined as
		\begin{equation}
		P_{\vec{\phi, d}}(i,j) = \frac{n_{\phi,d}(i,j)}{N},
		\label{eq:}
		\end{equation}
		where $N$ is the total number of object pixels. Since there is no consensus about the choice of angles $\phi$ and distance $d$, the calculations have 
		been done with a distance of one pixel and $8$ directions considering only the nearest neighbourhood of each pixel. The used set of 
		angles $\Omega = \{\phi_0, \ldots, \phi_7 \}$ is given by
		\begin{equation}
		\phi_l = l \cdot 45^{\circ{}} \,, \ l = 0, \ldots, 7 \,, \ \phi_l \in \Omega.
		\label{eq:}
		\end{equation}
		This set of angles provides 8 matrices $P_{\phi, d}(i,j)$, one matrix for each of the directions. The co-occurrence matrix $C(i,j)$ is finally defined as
		\begin{equation}
		C(i,j) = \mbox{avg} \left( P_{\phi, d}(i,j) \right).
		\label{eq:}
		\end{equation}
		The entry $(i,j)$ in the co-occurrence matrix $C(i,j)$ therefore indicates the probability for the co-occurrence of pixel value $i$ with 
		pixel value $j$ within the nearest neighbourhood. In Figure~\ref{fig:figure2} the corresponding co-occurrence matrix for a filament channel, shown 
		in Figure~\ref{fig:figure1}, is visualized. For the implementation it is necessary to differentiate between the actual pixel value and the 
		position in the co-occurrence matrix.
		
		Important properties of the co-occurrence matrix $C(i,j)$ are represented in Figure~\ref{fig:figure2}: (i) $C(i,j)$ is always a symmetric matrix since 
		the number of occurrences for pixel value $i$ together with pixel value $j$ will be always the same as vice versa. (ii) $C(i,j)$ provides a probability 
		density function (PDF), to find probabilities of joint relationships between a pair of pixel values $i$ and $j$. Thus, the sum over all entries 
		must be equal to $1$.

		\subsubsection{Textural Features}
		A set of textural features (Appendix~\ref{sec:texturalfeatures}) can be calculated from the co-occurrence matrix $C(i,j)$.
		For illustrative propose, we discuss two textural features (energy $H_1$ and contrast $H_2$) in detail:
		\begin{equation}
		H_1 = \sum_i{\sum_j{C(i,j)^2}},
		\label{eq:}
		\end{equation}
		\begin{equation}
		H_2 = \sum_i{\sum_j{(i - j)^2 \ C(i,j)}}.
		\label{eq:}
		\end{equation}
		Energy $H_1$ is a measure of homogeneity of the pixel pair distribution. It is high for a small number of different pixel pair combinations and low 
		for a large number of different combinations. $H_1$ is at its maximum if the image contains only a single pixel value, independent of the pixel 
		value itself. Thus, the energy is mainly a measure of the sharpness of the peak in the co-occurrence matrix. It is high for a sharp peak, indicating 
		only a few pixel value combinations, and low for a broad peak, indicating a variety of pixel value combinations. 
		
\begin{figure}
			 \centering
			 \includegraphics[width=0.55\columnwidth]{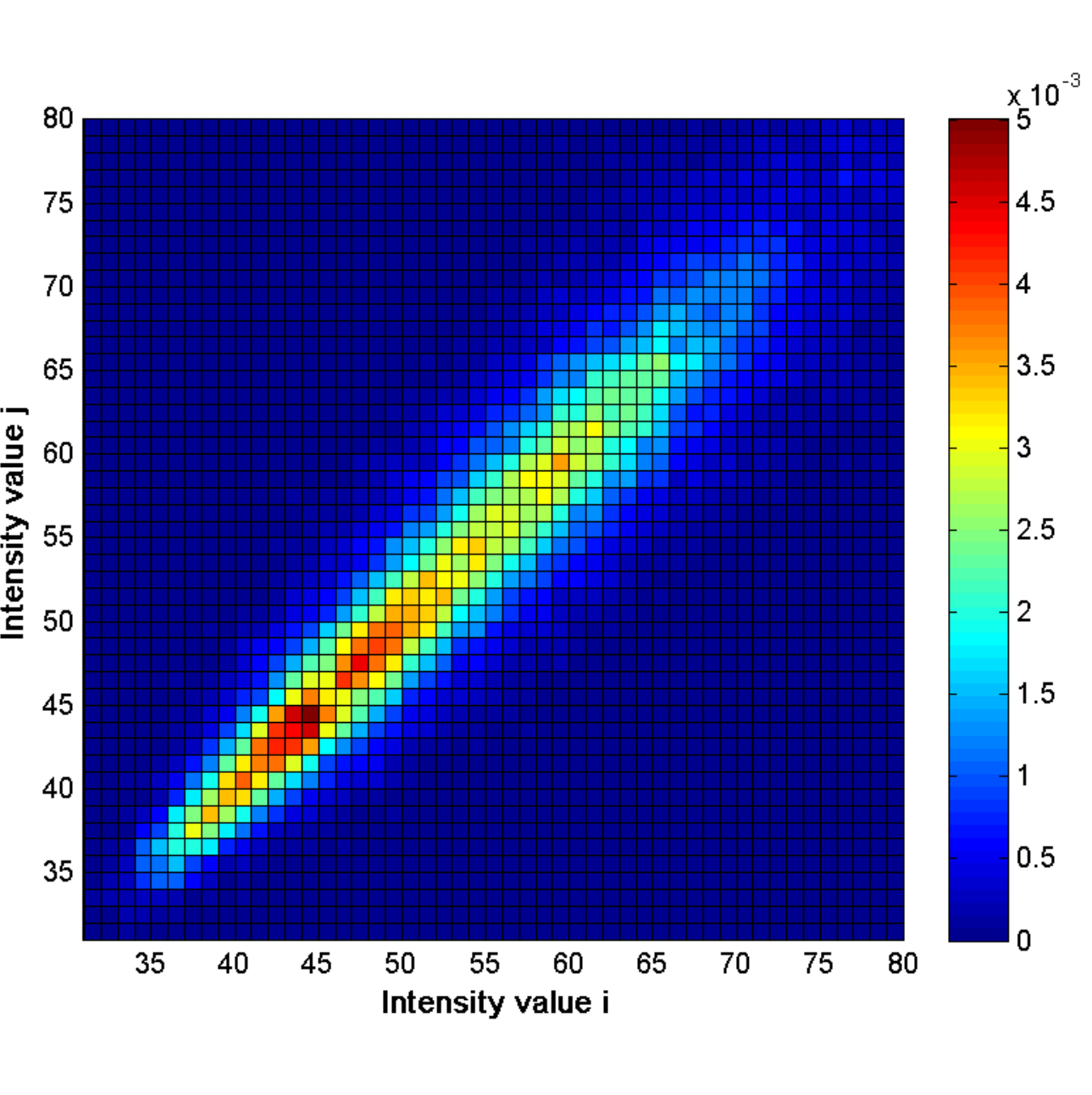}
			 \caption{\small Representation of the co-occurrence matrix $C(i,j)$ calculated from the filament channel shown in Figure~\ref{fig:figure1}. The
			colorbar of the co-occurrence matrix entries $C(i,j)$ represent the probability for neighbouring occurrences of pixel value $i$ with pixel value $j$.}
			 \label{fig:figure2}
		\end{figure}

		Due to the quadratic influence of the differences in the contrast $H_2$, the value is high for large differences between pixel value $i$ and 
		pixel value $j$. Hence, mainly the position of the peak in the co-occurrence matrix is important (see Figure~\ref{fig:figure2}). The position indicates small or 
		large differences of the pixel values. The contrast feature is a measure of the amount of local variations present in the image. It is high for large 
		differences of pixel pair values and low for small differences. All textural features that were used in this analysis are listed 
		in Appendix~\ref{sec:texturalfeatures}. 

		\section{Supervised classification problem}\label{S:classificationproblem} 
		The computed attributes were used as input for data mining in order to design the most suitable decision rule for a separation between coronal holes and filaments.  
		Supervised classification is commonly applied when a large set of attributes is available and the interpretation of the obtained information is complicated. 
		In supervised classification problems, an object is observed and we aim to classify it into one of two classes ($0$ or $1$, here \lq filament' or \lq coronal hole'). 
		To make this decision we have access to measurements of various properties of the object, called \emph{features}. Given these feature vectors $\mathbf{x}\in R^d$, 
		we aim to find a decision (or classification) rule, that is, a mapping $c: R^d \rightarrow \{0,1\},$ where $c(\mathbf{x})$ indicates the decision when feature 
		vector $\mathbf{x}$ is observed. Out of the numerous collection of possible decision rules, we aim to identify those performing best. Four widely used algorithms 
		with proven theoretical properties (Section~\ref{S:algo}) are evaluated via a common protocol (Section~\ref{S:evaluation}). Section~\ref{sec:results} describes the 
		performance measures used to compare them. 

		The set of attributes described in Section~\ref{sec:shapemeasures}-\ref{sec:2ndorderstatistics} together with the labelling of the maps as 
		\lq filament' or \lq coronal hole' provide us with a labeled dataset of feature values. We consider two situations. First, the whole set of attributes, 
		computed with both SDO/AIA and SDO/HMI from SPoCA, is used as feature values with the aim to reach the best classification. Second, we create new classifiers trained on a sub-set of attributes using only AIA information on SPoCA maps. The aim is twofold: (i) obtain a set of rules that are easy to implement into the SDO EDS pipeline, (ii) measure the corresponding improvement in performance for the SPoCA algorithm.

				\subsection{Supervised classification algorithms}\label{S:algo}
				We tested four classifiers using the scikit-learn library~\citep{scikit-learn}:
				\begin{description}
				\item[\bf{Linear Support Vector Machine}]

				The key idea of Linear Support Vector Machine (Linear SVM) is to find hyperplanes that separate the data as much as possible, that is, with a 
				large margin. SVM optimizes a trade-off between maximizing the margin of separability between classes and minimizing margin errors. 
				It provides a convex approximation to the combinatorial problem of minimizing classification errors. The practical implementation is 
				formulated as the minimization of a penalized loss function. We used the LIBLINEAR~\citep{liblinear} implementation of Linear SVM.

				\item[\bf{Support Vector Machine}]

				A second key idea of SVM as presented by Vapnick in his original formulation~\citep{svm} is to map the feature vectors in a nonlinear way 
				to a high (possibly infinite) dimensional space and then utilize linear classifiers in this new space. This is done through the use of a 
				kernel function. In our case, we tried several kernel functions: gaussian, sigmoid, polynomial and linear from the LIBSVM library~\citep{libsvm}.

				\item[\bf{Decision Tree}]
				
				Decision trees produce a set of  if-then-else decision rules that are selected on the basis  of the expected reduction in entropy that 
				would result from sorting a particular attribute. Trees are usually grown to their maximum size before a pruning step is applied to reduce 
				over-fitting. Various decision tree algorithms have been proposed~\citep{breiman,c45}. For this work we used decision trees using the 
				Gini and entropy criteria and various depths and splitting rules\footnote{\url{http://scikit-learn.org/stable/modules/tree.html}}.

				\item[\bf{Random Forest}]
				
				We tested an ensemble classifier method\footnote{\url{http://scikit-learn.org/stable/modules/ensemble.html}}. More precisely, we used 
				Random Forest, where a set of decision trees is created by introducing some randomness in the construction of the decision tree. 
				The prediction of the ensemble is given as the averaged prediction of the individual decision tree~\citep{random_forest}.

				\end{description}

				\subsection{Training and evaluation protocol}
				\label{S:evaluation}
				The supervised classification in machine learning is always performed in two steps. 1.) During the training phase, a model is estimated 
				from the data. 2.) In the validation (or test-) phase the trained model is applied on other data and model properties (such as error 
				classification rate) are estimated. This is done in practice by splitting the initial data set into a \lq training dataset' and a 
				\lq validation dataset'. 

				However, in the present study we want to compare the performance of several classifiers. In this case, performing the splitting only once is 
				not enough, as the observed difference between two classifiers may depend on the chosen training and test samples. To avoid this, 
				we need to repeat the comparison over randomly selected partitions of the data and report the average performance (see Chap.~5 in \cite{evaluatingLearning}
				for a discussion on error estimation in classification problem).

				We therefore perform 100 iterations of the following protocol. We do a stratified shuffle-split of the original dataset into a 75\% 
				development set and 25\% evaluation set. This means that we shuffle the original dataset, then split it into two 75\% / 25\% subsets (shuffle-split) 
				where each subset has approximately the same class distributions as the full dataset (stratification). The development set in each iteration is 
				used to train and evaluate each hyper-parameter combination for each algorithm. To choose the best hyper-parameter combination we use 
				stratified 5-fold cross-validation. A $k$-fold cross-validation means that the development set is further split into $k$-folds. 
				Each combination of $(k-1)$ folds is used for training and the remaining fold serves for testing, for a total of $k$ train/test splits. The use 
				of a \emph{stratified} 5-fold cross-validation means that each fold has approximately the same class distributions as the original dataset.

				Once the optimum combination of hyper-parameters is found, it is used to train a classifier on the 75\% development set, and is evaluated on 
				the 25\% evaluation set. This final hold-out evaluation set is necessary to accurately estimate real-world 
				performance because the cross-validation can over-fit for the particular split. The performance is measured by computing a skill score for each 
				of the 100 iterations, see Section~\ref{sec:results}. This allows us to quantify an average performance, but also to evaluate the variance in 
				performance results across different runs. The analysis code can be accessed online\footnote{\url{https://github.com/rubendv/ch_filament_classification}}.

\section{Results}\label{sec:results}

		To evaluate the performance of the four classifiers we computed for each of the 100 runs the \lq confusion matrix'. A confusion matrix, see 
		Table~\ref{T:confusion}, contains the elements TP (true positive, coronal hole predicted and observed), FP (false positive, coronal hole predicted 
		and filament channel observed), FN (false negative, filament channel predicted and coronal hole observed) and TN (true negative, filament channel 
		predicted and filament channel observed). 
		
		\begin{table}[h]
		\begin{center}
		\begin{tabular}{@{}lll@{}}
		\toprule
		Predicted & Observed: Coronal Hole (CH) & Filament Channel (FC) \\ \midrule
		CH        & True Positive (TP)          & False Positive (FP)   \\
		FC        & False Negative (FN)         & True Negative (TN)    \\ \bottomrule
		\end{tabular}
		\end{center}
		\caption[]{Coronal hole and filament channel classification contingency table (confusion matrix). \label{T:confusion}}
		\end{table}

		There exists a variety of skill scores~\citep{woodcock76} built by combination of the confusion matrix elements. One of the most frequently used is the Hanssen \& Kuipers 
		discriminant also known as true skill statistics (TSS) \citep{hanssen65, bloomfield12}. The TSS is defined as the proportion of correctly predicted coronal holes among all 
		coronal holes (TPR) minus the proportion of filaments that were classified as coronal holes among all filaments (FPR):\footnote{For the reader 
		who is used to the terms \lq recall' and  \lq precision' we note that recall is equivalent to the TPR, and precision denotes the fraction of correctly predicted coronal holes among 
		all predicted coronal holes. The TSS is given by $\mbox{TP/(TP+FN)} + \mbox{TN/(TN+FP)} - 1$, the sum of recalls for coronal holes and filament channels minus a scaling factor of 1.}
		
		\begin{equation}
		\mbox{TSS} = \mbox{TPR} - \mbox{FPR} = \frac{\mbox{TP}}{\mbox{TP+FN}} - \frac{\mbox{FP}}{\mbox{FP+TN}}.
		\label{eq:}
		\end{equation}

		\noindent The TSS is defined in the range $[-1,1]$. A TSS of $0$ indicates that the algorithm cannot distinguish between coronal holes and 
		filament channels. A perfect classifier would have the value $1$ or $-1$ (inverse classification), respectively. In~\cite{bloomfield12} the TSS was proposed to be the standard skill 
		score for comparing the performances of flare forecasts with differing flare/no-flare sample ratios and is also appropriate for this study. When applying the developed algorithms 
		in the form of an automated coronal hole detection tool we are mainly interested in both, the proportion of correctly predicted coronal holes (TPR) as well as the proportion of falsely 
		predicted coronal holes (FPR). Hence, the combination of TPR and FPR in form of the TSS provides an important and intuitive indicator for the performance of the developed framework.
		Furthermore, the TSS uses all elements of the confusion matrix and is the only known skill score which is unbiased by the proportion of coronal holes and filament channels in the 
		data sets.

\begin{figure}%
		\centering
		\includegraphics[width=.95\linewidth]{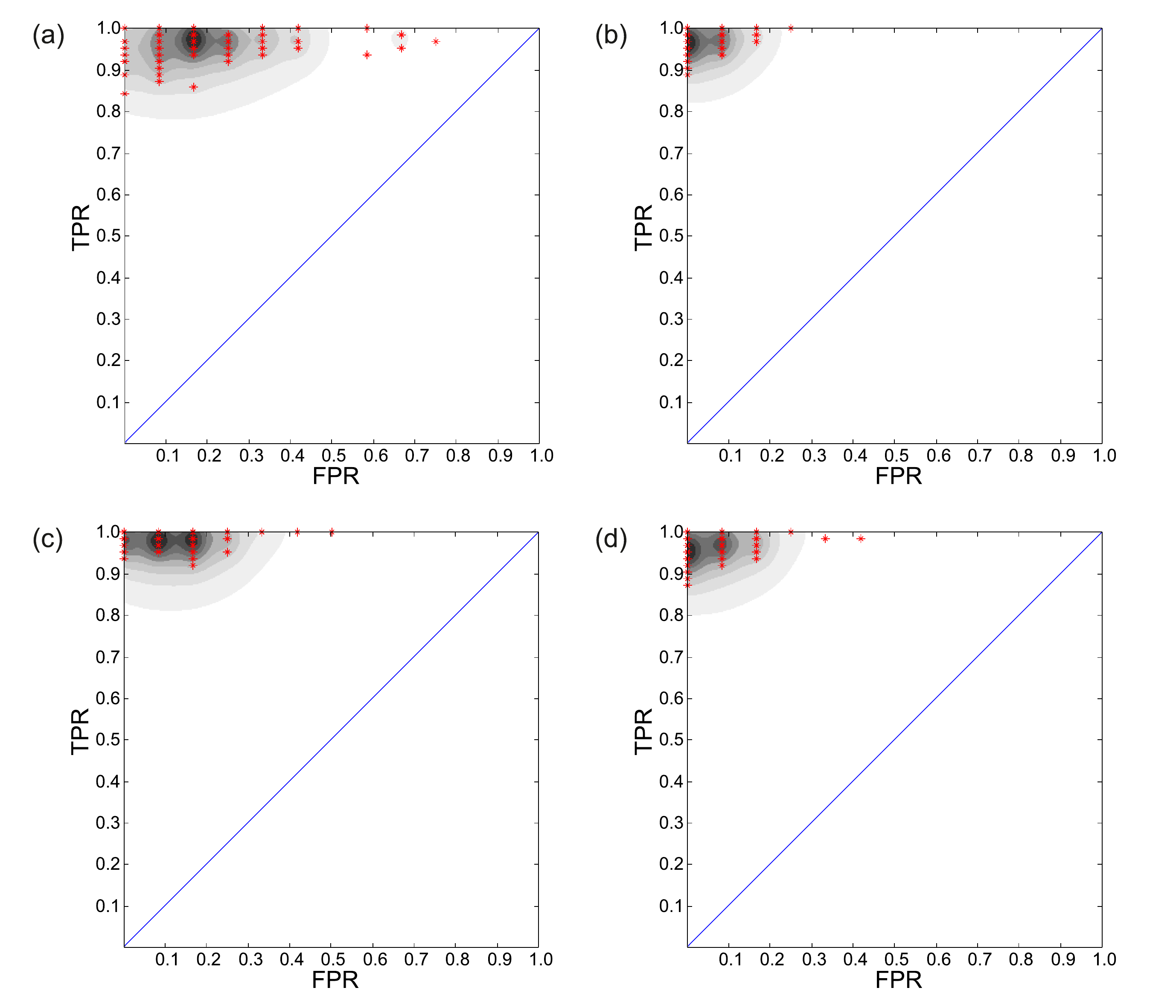}
		\caption{Density plot of correctly predicted coronal holes (TPR) versus erroneously predicted coronal holes (FPR) obtained from intensity-based coronal hole 
		region segmentation using all attributes. The red crosses denote the individual results for the TPR and FPR for each of the 100 iterations, according
		to the training and evaluation protocol described in Section~\ref{S:evaluation}. (a) Decision Tree; (b) Linear SVM; (c) Random Forest; (d) SVM.}
		\label{fig:figure3}%
		\end{figure}

\begin{figure}%
		\centering
		\includegraphics[width=.95\linewidth]{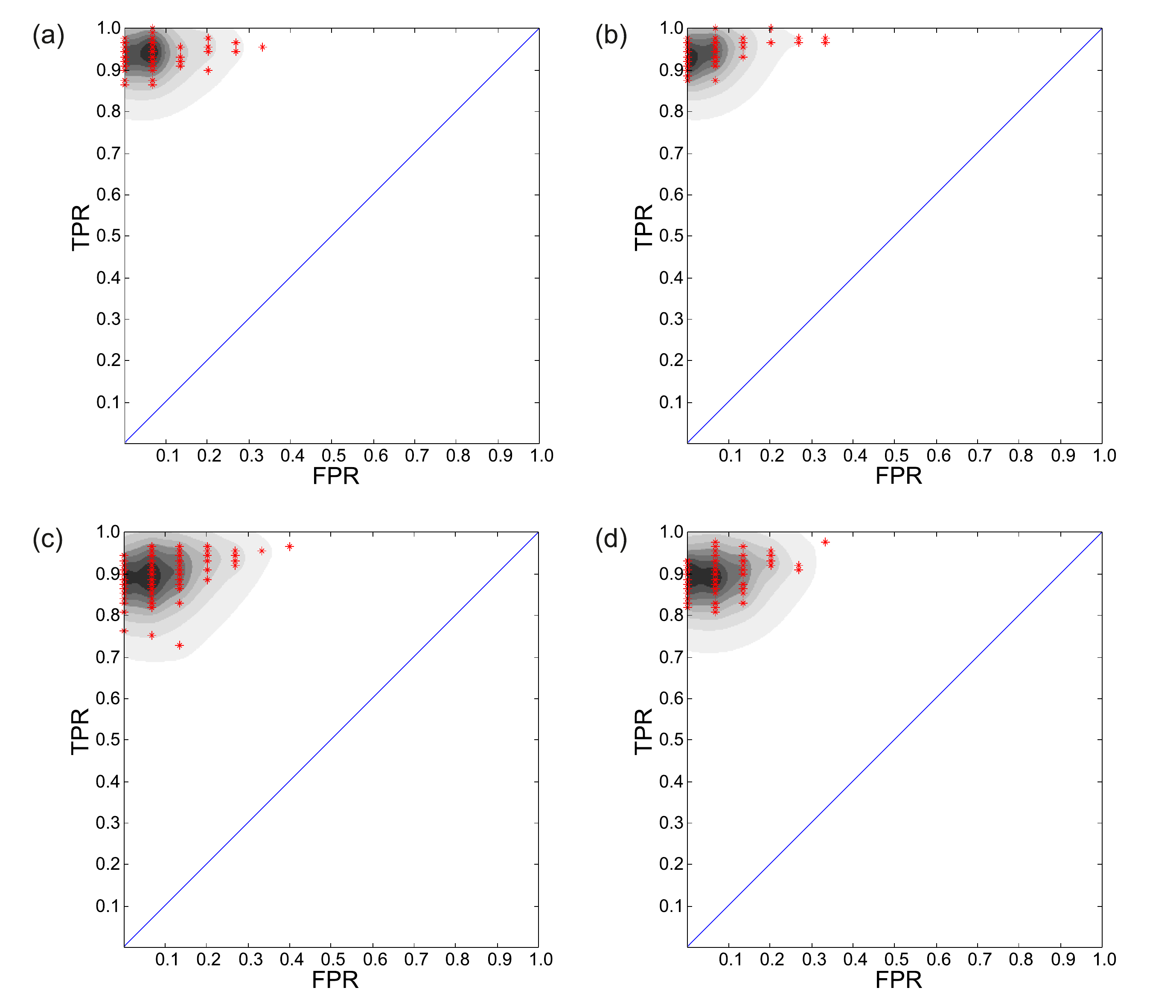}
		\caption{Density plot of correctly predicted coronal holes (TPR) versus erroneously predicted coronal holes (FPR) obtained with SPoCA algorithm. The red crosses denote the individual results for the TPR and FPR for each of the 100 iterations, according to the training and evaluation protocol described in Section~\ref{S:evaluation}. (a) SVM (all attributes); (b) Linear SVM (all attributes); (c) SVM (AIA attributes only); (d) Linear SVM (AIA attributes only).}
		\label{fig:figure4}%
		\end{figure}

		Figure~\ref{fig:figure3} (resp. \ref{fig:figure4}) shows for the intensity based threshold method (resp. for the SPoCA method) the proportion of correctly predicted coronal 
		holes (TPR) (that is, the efficiency or sensitivity) versus the proportion of falsely predicted coronal holes (FPR). The performance for each of the 100 runs is visually 
		represented in those figures by the points $(\mbox{FPR}, \mbox{TPR})$. A perfect classification would have only points at the position $(0,1)$, a perfect inverse classification 
		points at $(1,0)$ and \lq no performance' is represented by the diagonal line ($\mbox{FPR} = \mbox{TPR}$). 
	
\begin{figure}%
		\centering
		\includegraphics[width=.75\linewidth]{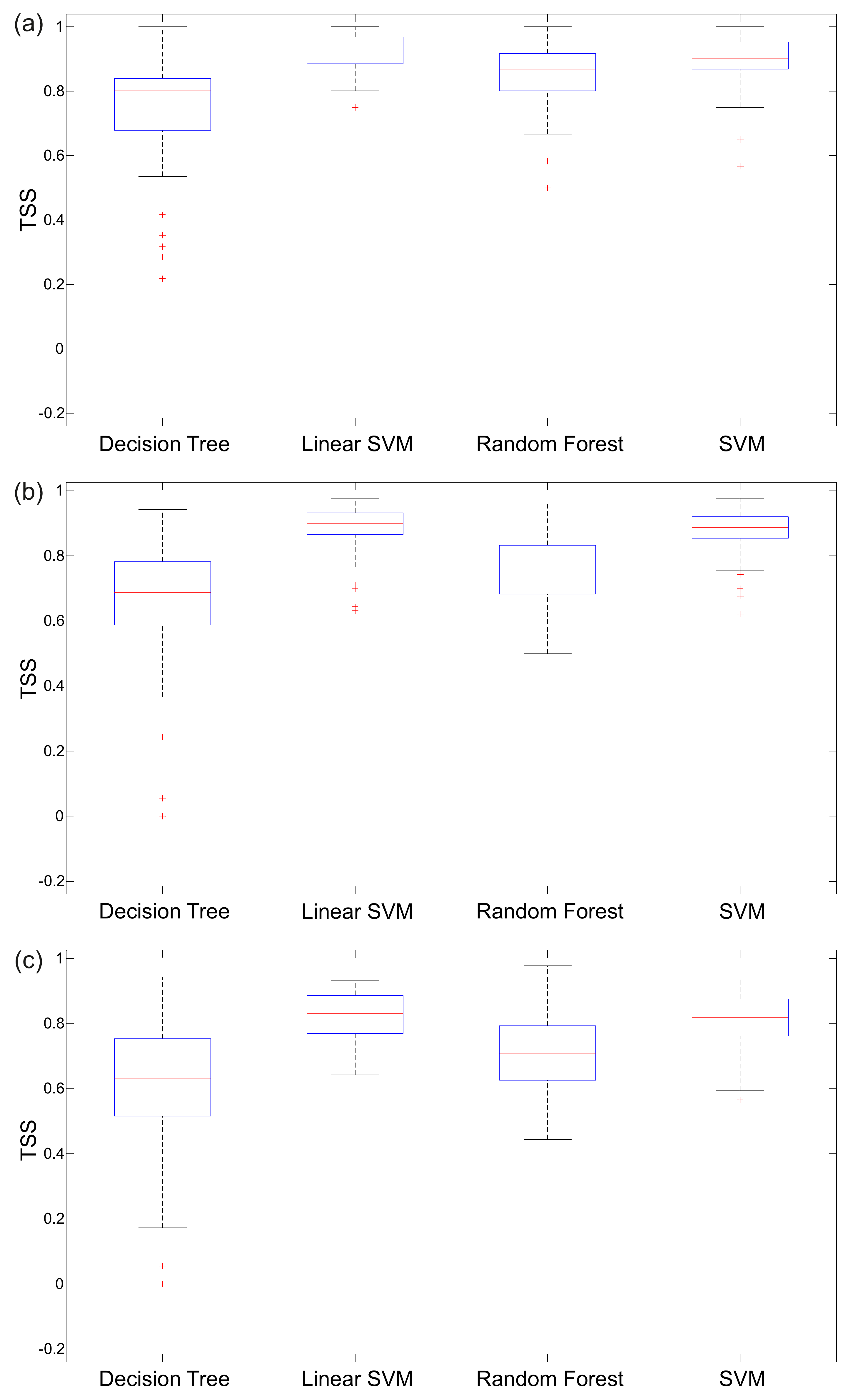}
		\caption{Comparison between the computed true skill statistics (TSS) for different classifiers using different segmentation techniques and sets of attributes. On each 
		box, the central mark is the median ($50$th percentile), the edges of the box are the $25$th and $75$th percentiles, the whiskers show the $\pm 2.7 \sigma$ range covering 
		$99.3\%$ of the data and outliers are plotted individually as red crosses. (a) Intensity Threshold (all attributes) (b) SPoCA (all attributes); 
		(c) SPoCA (only attributes from AIA).}%
		\label{fig:boxplots}%
		\end{figure}

		Figure~\ref{fig:figure3} shows the performance of the four classifiers using all attributes (from AIA and magnetograms) computed with the intensity-based threshold 
		segmentation maps. All classifiers produce a high TPR, but only the two SVMs achieve at the same time a consistently low FPR, with Linear SVM performing 
		slightly better than SVM. In Figure~\ref{fig:figure4} the results for the SPoCA detection of SVM and Linear SVM are presented. Both classifiers achieve high 
		TPR and low FPR. A comparison of the performance with and without magnetogram information indicates that the inclusion of magnetogram-based attributes in addition to the AIA-based attributes systematically improves both TPR and FPR measures across all four classifiers: Figure~\ref{fig:figure4} compares the density of TPR versus FPR obtained for both SVM classifiers when adding or not the magnetogram information, Table~\ref{tab:table2} gives the median TSS and the standard deviation for all tested classifiers, and Figure~\ref{fig:boxplots} provides a graphical comparison of the actual TSS values in the form of box plots.

		\begin{table}[h]
		\begin{center}
		\begin{tabular}{@{}llcccc@{}}
		\toprule
		\multicolumn{2}{c}{}                    & \multicolumn{4}{c}{\textbf{True Skill Statistics (TSS)}}                               \\ \cmidrule(l){3-6} 
		\multicolumn{2}{c}{\textbf{Datasets}}   & \textit{SVM}  & \textit{Linear SVM} & \textit{Random Forest} & \textit{Decision Tree} \\ \midrule \midrule
		\multicolumn{2}{l}{Intensity Threshold (all attributes)} & $0.90 \pm 0.07$ & $0.94 \pm 0.05$       & $0.87 \pm 0.10$          & $0.80 \pm 0.15$           \\
		\multicolumn{2}{l}{SPoCA (all attributes)}    & $0.89 \pm 0.07$ & $0.90 \pm 0.06$       & $0.77 \pm 0.11$          & $0.68 \pm 0.18$           \\
		\multicolumn{2}{l}{SPoCA (only attributes from AIA)}           & $0.82 \pm 0.08$ & $0.83 \pm 0.07$       & $0.71 \pm 0.11$          & $0.63 \pm 0.21$           \\ \bottomrule
		\end{tabular}
		\end{center}
		\caption[]{Median TSS and standard deviation for all four classifiers and the three different datasets.}
		\label{tab:table2}%
		\end{table}
				
		For both segmentation techniques best results are achieved with linear SVM showing the highest median TSS and smallest standard deviation. Similar results can be 
		obtained with SVM, although with slightly lower values of the median TSS. In contrast, Decision Tree and Random Forest exhibit significantly lower median TSS and higher 
		standard deviation. For the dataset extracted with the SPoCA algorithm we also considered the case where attributes are obtained from AIA images only. The corresponding 
		performance is, as expected, lower than when magnetogram information is included. 
		
		Another comparison of the investigated classifiers is presented by the Receiver Operator Characteristic (ROC) curves. ROC curves 
		are commonly used to compare and evaluate the performance of different classifiers. They characterize how the number of correctly classified coronal holes (TPR) 
		varies with the number of incorrectly classified filament channels (FPR). This gives an illustration of the tradeoff between the completeness of coronal holes and the 
		contamination with filaments for each classifier. Hence, the ROC curves convey the information of which classifier should be applied when either a small, focused sample 
		of coronal holes or a larger sample of coronal holes is requested. 
		
\begin{figure}%
		\centering
		\includegraphics[width=.9\linewidth]{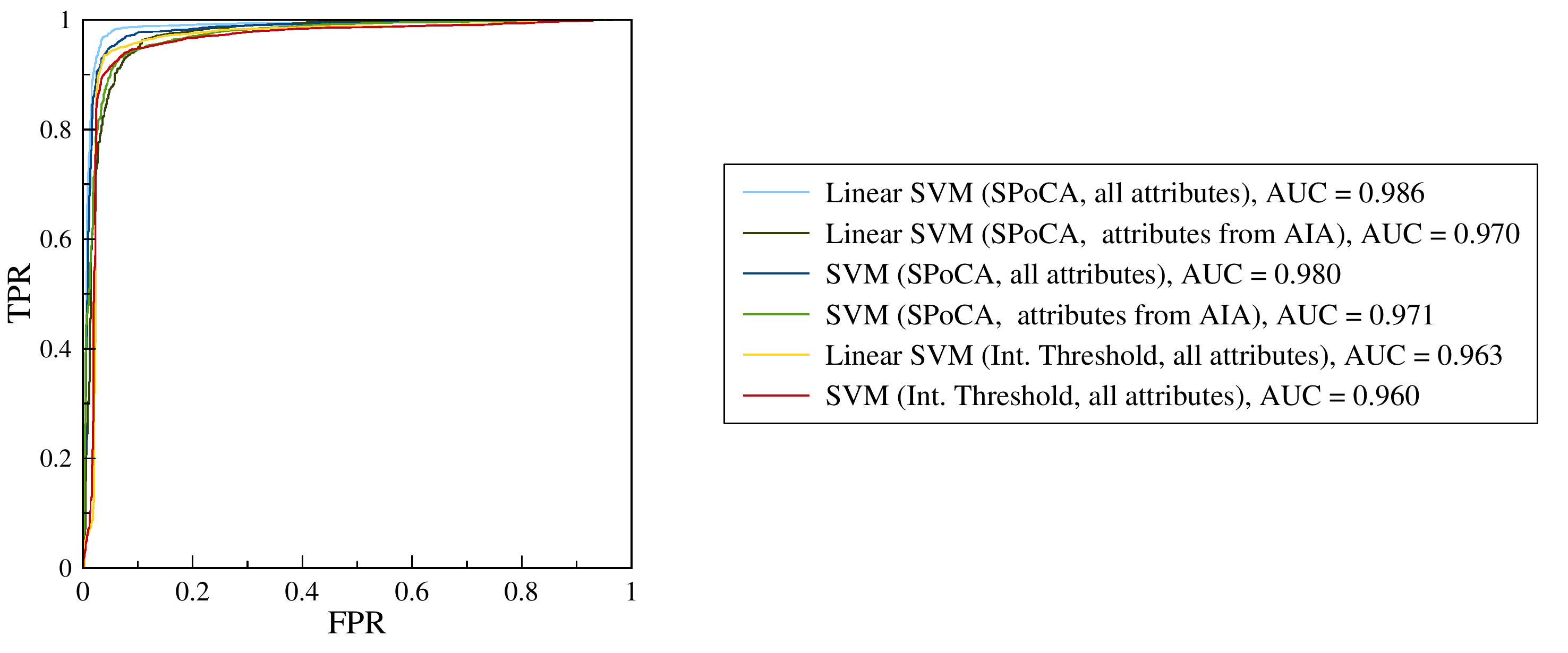}
		\caption{ROC curves for the SVM and Linear SVM classifier together with the calculated area under the curve (AUC).}%
		\label{fig:roc_curves}%
		\end{figure}

		In Figure~\ref{fig:roc_curves}, the ROC curves for SVM and linear SVM calculated from the combined results of 
		each of the 100 iterations are shown. Again, the FPR is plotted on the x-axis and the TPR on the y-axis. The best results are represented by curves close to the upper left 
		corner with a possible large area under the curve (AUC). The AUC is usually used in order to roughly summarize the obtained ROC curves in a single quantity. The best 
		performances, indicated by a high completeness together with a low contamination, are achieved with the Linear SVM followed by the SVM using all computed attributes. 
		The ROC analysis confirms that the exclusion of the magnetogram information negatively affects the performances of both classifiers by decreasing the TPR and increasing the FPR.
		
\section{Discussion and Conclusion}\label{sec:discussion}

	The most popular methods for the extraction of coronal hole regions are based on the intensity information in EUV images of the Sun. However, other features, 
	notably filament channels, appear similarly dark as coronal holes. It is a general problem of intensity-value-based techniques to distinguish between those features. 
	We found in our carefully prepared datasets using two different segmentation algorithms (intensity-based thresholding and SPOCA) that about $15\%$ of the coronal 
	hole candidates are actually filament channels. We therefore developed new algorithms and techniques to tackle this shortcoming.
	
	We investigated a new approach using a combination of segmentation techniques and machine learning algorithms. The proposed attributes were 
	used as input for supervised classification schemes. Four commonly applied classifiers, including SVM, Linear SVM, Decision Tree and Random Forest were tested for two 
	different segmentation techniques and evaluated based on true skill statistics. The results reveal that all classifiers provide good results in general, with the linear SVM 
	and SVM classifier providing the best performances (TSS $\approx 0.90$). We also found that the two different image segmentation techniques provide very similar results 
	indicating the broad applicability of the presented method. In addition to EUV images we used the magnetic field information from HMI line-of-sight magnetograms. Adding the information and extracted attributes from the magnetic field data improves the performance across all four classifiers for the SPoCA detection (c.f. Table~\ref{tab:table2}).
	
	Besides standard statistical attributes we investigated the benefits from textural features, originally introduced by \cite{haralick73}, to analyze the texture 
	information of coronal holes and filament channels in high spatial resolution AIA 19.3~nm and line-of-sight magnetograms. For this purpose we adapted the original 
	method introduced by Haralick to open value ranges in order to calculate a set of textural features for intensity and magnetic field configurations. This enabled 
	us to describe the intrinsic spatial arrangement contained within coronal hole and filament channel regions in AIA 19.3~nm and line-of-sight magnetograms with a set 
	of textural features. Textural features provide information about the structural arrangement of pixel values in contrast to first order statistics, which focus on the 
	overall pixel value distribution contained in an image. In this respect we note that for the detailed analysis of textural features one has to take into account the influence of noise. 
	Digital images are discrete representations of objects or scenes, unavoidably, they have limited spatial resolution and they are contaminated with noise \citep{delouille08}. 
	Hence, the image quality is a limiting factor for using such kind of methods and low signal to noise ratios may hinder the application of textural information.

	The algorithms are not costly with respect to computing-time. The segmentation algorithms as well as the calculation of attributes and classifiers are calculated within 
	a few minutes. This is favorable for real-time application. We therefore aim to implement the described algorithms in currently available forecasting tools for solar wind 
	high speed streams\footnote{e.g., \url{swe.uni-graz.at/solarwind} run by UNIGRAZ}. We also plan on improving the SPoCA-CH module within the SDO EDS pipeline using Linear 
	SVM classification on AIA attributes. 
	
	In summary, this study demonstrates how a machine learning approach may help improve upon an unsupervised feature extraction method\footnote{The machine learning code used for the present analysis is available at \url{https://github.com/rubendv/ch_filament_classification}}. The findings suggest that the proposed 
	attributes and classifiers may actually be applicable for a wide range of imaging data. We stress that the inclusion of magnetic field data improved the 
	performance of the coronal hole and filament discrimination in EUV images. All tested classifiers provide good results for the 
	TSS and we therefore expect that the developed detection tool has the potential to reduce coronal hole classification error rate for both segmentation algorithms.
	
\begin{acknowledgements} 
	We gratefully acknowledge the NAWI Graz funding \textit{F\"orderung von JungforscherInnengruppen 2013--15}. The research leading to these results
	has received funding from the European Commission's Seventh Framework Programme (FP7/2007-2013) under the grant agreement no.\,284461 [eHEROES]. 
	R.\,DV. and V.\,D. acknowledge support from the Belgian Federal Science Policy Office through the BRAIN.be and the ESA-PRODEX programs. M.\,T. acknowledges 
	the Fonds zur F\"orderung wissenschaftlicher Forschung (FWF): V195-N16. M.A.\,R. acknowledges support from the association \textit{Dynamics of the Solar System}  
	to finance a one-month stay at the Royal Observatory of Belgium. We thank P.~Dupont, N.~Sabathiel and T.~Rotter for insightful discussions. The editor thanks Shaun Bloomfield and Jake VanderPlas for their assistance in evaluating this paper.
\end{acknowledgements}






\begin{appendix}
\section{Textural Features}
\label{sec:texturalfeatures}
		\subsection{Notation}
				We suggest to use a set of textural features which can be calculated from the co-occurrence matrix. 
				The following equations define these features.
				\[
				\begin{array}{lp{0.8\linewidth}}
					 C(i,j) & $(i,j)$-\textit{th} entry in the normalized co-occurrence matrix. \\
					 p_x(i) & $i$-\textit{th} entry is obtained by summing the rows of $C(i,j)$. \\
					 p_y(j) & $j$-\textit{th} entry is obtained by summing the columns of $C(i,j)$. \\
					 p_{x-y}(k) \ \ & $k$-\textit{th} entry of $p_{x-y}(k)$ corresponding to the sum over all entries of $C(i,j)$ with absolute 
					pixel value difference of $i$ and $j$ equal to $k$.\\
					 p_{x+y}(k) & $k$-\textit{th} entry of $p_{x+y}(k)$ corresponding to the sum over all entries of $C(i,j)$ where
					the addition of $i$ and $j$ is equal to $k$.\\
					 \mu_x, \mu_y & means of $p_x$ and $p_y$. \\
					 \sigma_x, \sigma_y & standard deviations of $p_x$ and $p_y$. \\
					 N_g & Number of distinct pixel values. \\
					 \sum_i , \sum_j & Convention indicating $\sum_{i=1}^{N_g}$, $\sum_{j=1}^{N_g}$.\\
					 $\mbox{HX, HY}$ & Entropies of $p_x$ and $p_y$
				\end{array}
				\]
				\begin{equation}
				p_x(i)=\sum_j{C(i,j)}, \, , \ p_y(j)=\sum_i{C(i,j)},
				\end{equation}
				\begin{equation}
				p_{x+y}(k)=\sum_i{\sum_j{C(i,j)}} \,, \ i+j=k \,, \ k=2,3, \ldots, 2 N_g,
				\end{equation}
				\begin{equation}
				p_{x-y}(k)=\sum_i{\sum_j{C(i,j)}} \,, \ |i-j|=k \,, \ k=0,1, \ldots, N_g - 1,
				\end{equation}
				\begin{equation}
				\mbox{HX} = - \sum_i{p_x(i)\ \log(p_x(i))} \, , \ \mbox{HY} = - \sum_i{p_y(i)\ \log(p_y(i))}
				\end{equation}
				\begin{equation}
				\mbox{HXY1} = - \sum_i{\sum_j{C(i,j) \log(p_x(i) p_y(j))}},
				\label{eq:}
				\end{equation}
				\begin{equation}
				\mbox{HXY2} = - \sum_i{\sum_j{p_x(i) \ p_y(j) \log(p_x(i) \ p_y(j))}},
				\label{eq:}
				\end{equation}

		\subsection{Textural Features}
				Based on this notation, the following textural features can be calculated: 
				\\
				\\
				Energy:
				\begin{equation}
				H_1 = \sum_i{\sum_j{C(i,j)^2}},
				\label{eq:}
				\end{equation}
				Contrast:
				\begin{equation}
				H_2 = \sum_i{\sum_j{(i - j)^2 C(i,j)}},
				\label{eq:}
				\end{equation}
				Correlation:
				\begin{equation}
				H_3 = \sum_i{\sum_j{\frac{(i - \mu_x)(j - \mu_y)}{\sigma_x \sigma_y}}},
				\label{eq:}
				\end{equation}
				Sum of Squares - Variance:
				\begin{equation}
				H_4 = \sum_i{\sum_j{(i-\mu)^2 C(i,j)}},
				\label{eq:}
				\end{equation}
				Homogeneity:
				\begin{equation}
				H_5 = \sum_i{\sum_j{\frac{C(i,j)}{1 + (i - j)^2}}},
				\label{eq:}
				\end{equation}
				Sum Average:
				\begin{equation}
				H_6 = \sum_{i=2}^{2 N_g}{i \ p_{x+y}(i)},
				\label{eq:}
				\end{equation}
				Sum Variance:
				\begin{equation}
				H_7 = \sum_{i=2}^{2 N_g}{(i-H_8)^2 p_{x+y}(i)},
				\label{eq:}
				\end{equation}
				Sum Entropy:
				\begin{equation}
				H_8 = -\sum_{i=2}^{2 N_g}{p_{x+y}(i) \log{(p_{x+y}\left(i\right))}},
				\label{eq:}
				\end{equation}
				Entropy:
				\begin{equation}
				H_9 = -\sum_i{\sum_j{C(i,j) \log{(C(i,j))}}},
				\label{eq:}
				\end{equation}
				Difference Variance: 
				\begin{equation}
				H_{10} = var\left( p_{x-y} \right),
				\label{eq:}
				\end{equation}
				Difference Entropy: 
				\begin{equation}
				H_{11} = -\sum_{i=0}^{N_g - 1}{p_{x-y}(i) \log(p_{x-y}(i))},
				\label{eq:}
				\end{equation}
				Information Measures of Correlation (I):
				\begin{equation}
				\frac{H_9 -\mbox{HXY1}}{\mbox{max} \mbox{(HX,HY)}}
				\label{eq:}
				\end{equation}
				Information Measures of Correlation (II):
				\begin{equation}
				H_{13} = \sqrt{ 1- \exp (-2 \ ( \mbox{HXY2} - H_9) \ ) }.
				\label{eq:}
				\end{equation}
\end{appendix}
\end{document}